\documentstyle[a4wide,12pt]{article}
\title{A geometric formulation of strong interaction}
\author{Vu B Ho\\Department of Physics\\Monash University\\
Clayton Victoria 3168\\Australia}
\begin{document}
\maketitle

\begin{abstract}
It is observed that, at short range, the field equations of general
relativity admit a line element that takes the form of Yukawa potential.
The result leads to the possibility that strong interaction may also be
described by field equations that have the same form as that of general
relativity. It is then shown how such field equations may arise from the
coupling of two strong fields.
\end{abstract}

\newpage
\noindent
{\bf 1. Introductory remarks}\\
\\
Like the gravitational force, the strong force does not seem to relate to
the charge of a particle. Within the framework of classical physics, the
result leads to the suggestion of a strong field that must have a different
nature to the familiar fields of gravitation and electromagnetism. However,
since strong interaction has been investigated almost entirely in terms of
the quantum theory, the problems of what might be the classical dynamics of
a particle under the influence of strong force and what might be the
classical law of strong force and, in particular, what kind of matter that
produces the strong force have not been paid much attention. Obviously, the
most fundamental problem is whether the strong interaction has a classical
version. The problem could only be answered in a definite manner if it was
possible to investigate experimentally and classically the dynamics of
a particle moving in a strong force field.

In classical physics, however, since a particle is characterised only by
the mass and the charge, and the dynamics of the particle is assumed to
follow either the laws of classical electrodynamics or the laws of general
relativity, if the strong force does not have any kind of relationship with
the charge of a particle, then it would be possible to suggest the dynamics
of the particle would also follow the laws of general relativity, at least
in the form of the field equations. The present problem can be seen to be
similar to the problem of Newtonian gravitational field that forms a line
element that satisfies the field equations of general relativity. In the
case of strong force, perhaps, the most relevant form of potential that
would characterise a strong field would be that of Yukawa potential. Hence,
if the Yukawa potential also forms a line element solving the general
relativistic field equations, then there should be a close relationship
between strong interaction and the general formalism of relativity.\\
\\
\noindent
{\bf 2. A line element of Yukawa potential}\\
As mentioned above, in order to see whether the strong force should follow
the laws of general relativity, it is first to consider whether the Yukawa
potential can form a line element that satisfies the field equations of
general relativity

\begin{equation}
R_{\mu\nu}-\frac{1}{2}g_{\mu\nu}R=\kappa T_{\mu\nu}.
\end{equation}
Assuming a centrally symmetric field, the spacetime metric can be written

\begin{equation}
ds^2=e^\mu dt^2-e^\nu dr^2-r^2(d\theta^2+sin^2\theta d\phi^2).
\end{equation}
With this line element, the vacuum solutions satisfy the following system
of equations (Landau \& Lifshitz 1975)

\begin{eqnarray}
\frac{d\mu}{dr}+\frac{1}{r}-\frac{e^\nu}{r}&=&0,\\
\frac{d\nu}{dr}-\frac{1}{r}+\frac{e^\nu}{r}&=&0,\\
\frac{d\nu}{dt}&=&0,
\end{eqnarray}
\begin{equation}
2\frac{d^2\mu}{dr^2}+\left(\frac{d\mu}{dr}\right)^2+\frac{2}{r}\left
(\frac{d\mu}{dr}-\frac{d\nu}{dr}\right)-\frac{d\mu}{dr}\frac{d\nu}{dr} -
e^{\nu-\mu}\left(2\frac{d^2\nu}{dt^2}+\left(\frac{d\nu}{dt}\right)^2 -
\frac{d\nu}{dt}\frac{d\mu}{dt}\right)=0.
\end{equation}
These equations are not independent since it can be verified that the last
equation follows from the first three equations. Furthermore, the first two
equations give $d\nu/dr+d\mu/dr=0$ that also leads to $\nu+\mu=0$.

Now, if a line element in the form of Yukawa potential is assumed, then it is
obvious that it can be written in the following simple form

\begin{equation}
e^{-\nu}=1-\alpha\frac{e^{-\beta r}}{r}.
\end{equation}
By differentiation, it is found

\begin{equation}
\frac{d\nu}{dr}=-\alpha\frac{e^{-\beta r}}{r}\frac{1+\beta r}{r -
\alpha e^{-\beta r}}.
\end{equation}
On the other hand, the quantity $d\nu/dr$ can be deduced from the equation
(4) and the equation (7) as

\begin{equation}
\frac{d\nu}{dr}=-\alpha\frac{e^{-\beta r}}{r}\frac{1}{r-\alpha e^{-\beta r}}.
\end{equation}
The equation (8), by comparison, is reduced to the equation (9) if the
condition $\beta r\ll 1$ is satisfied. The quantity $\alpha$ plays the
role of the charge of a particle in electromagnetism, while the quantity
$\beta=1/R$, with $R=\hbar/mc$, specifies the range of strong force. The
quantity $m$ is the rest mass of virtual mesons whose continuous transfer
between two nucleons has been assumed to give rise to strong interactions.
The result showed that within the short range of strong force the field
equations of general relativity admit a line element that takes the
form of Yukawa potential.\\
\\
\noindent
{\bf 3. Strong interaction as a coupling of two strong fields}\\
\\
Since at short range it is possible to describe strong interaction by the
field equations of general relativity, the problem now is to look for what
might be the mechanism of the interaction. It is expected to be a spacetime
structure and to be formulated in terms of differential geometry. If the
affine connections assume the form $\Gamma^\sigma_{\mu\nu} =
\Lambda^\sigma_\mu\Phi_\nu$, then the curvature tensor

\begin{equation}
R_{\beta\mu\nu}^\alpha=\frac{\partial \Gamma_{\beta\nu}^\alpha}{\partial
x^\mu} - \frac{\partial \Gamma_{\beta\mu}^\alpha}{\partial x^\nu} +
\Gamma_{\lambda\mu}^\alpha \Gamma_{\beta\nu}^\lambda-
\Gamma_{\lambda\nu}^\alpha \Gamma_{\beta\mu}^\lambda
\end{equation}
can be shown to reduce to the form

\begin{equation}
R_{\beta\mu\nu}^\alpha = \frac{\partial {\left ( \Lambda_\beta^\alpha
\Phi_\nu \right )}}{\partial x^\mu} - \frac{\partial {\left (
\Lambda_\beta^\alpha \Phi_\mu \right )}}{\partial x^\nu},
\end{equation}
from which the Ricci tensor is found as (Ho 1994)

\begin{eqnarray}
R_{\mu\nu}&=&\frac{\partial{\left(\Lambda_\mu^\sigma \Phi_\nu\right)}}
{\partial x^\sigma}-\frac{\partial {\left ( \Lambda_\mu^\sigma
\Phi_\sigma \right)}} {\partial x^\nu}\\
&=& \left ( \frac{\partial \Phi_\nu}{\partial x^\sigma} - \frac{\partial
\Phi_\sigma}{\partial x^\nu} \right ) \Lambda_\mu^\sigma + \Phi_\nu
\frac{\partial \Lambda_\mu^\sigma}{\partial x^\sigma} -  \Phi_\sigma
\frac{\partial \Lambda_\mu^\sigma}{\partial x^\nu}.
\end{eqnarray}

Since the Ricci tensor that forms the field equations of general relativity
is symmetric, the above Ricci tensor needs first to be reduced to a
symmetric form. Assume the following relations between the quantities
$\Lambda^\sigma_\mu$ and $\Phi_\nu$

\begin{equation}
\Phi_\sigma\frac{\partial \Lambda^\sigma_\mu}{\partial x^\nu} =-
C^{\alpha\beta}_{\nu\sigma}\Phi_\alpha\Phi_\beta\Lambda^\sigma_\mu,
\end{equation}
where the quantities $C^{\alpha\beta}_{\nu\sigma}$ are arbitrary, but
supposed to be antisymmetric with respect to the subscript indices. This
leads to further assumption of the antisymmetry of the quantities
$\partial_\nu\Lambda^\sigma_\mu$ by imposing the conditions
$\partial_\nu\Lambda^\sigma_\mu=-\partial_\sigma\Lambda^\nu_\mu$. With
these requirements, the Ricci tensor then takes the form

\begin{eqnarray}
R_{\mu\nu}&=&\left(\frac{\partial \Phi_\nu}{\partial x^\sigma} -
\frac{\partial \Phi_\sigma}{\partial x^\nu} + C^{\alpha\beta}_{\nu\sigma}
\Phi_\alpha\Phi_\beta\right)\Lambda^\sigma_\mu\\
&=&F_{\sigma\nu}\Lambda^\sigma_\mu,
\end{eqnarray}
where the quantities $F_{\mu\nu}$ are defined by the relation

\begin{equation}
F_{\mu\nu}=\frac{\partial \Phi_\nu}{\partial x^\mu} -\frac{\partial
\Phi_\mu}{\partial x^\nu} + C^{\alpha\beta}_{\mu\nu}\Phi_\alpha\Phi_\beta.
\end{equation}
In this form it is seen that the quantities $F_{\mu\nu}$ can be regarded as
a strong field. Similar to the gravitational field, all strong sources of
the same characteristics are considered to produce identical strong fields,
then it is possible to identify the quantities $\Lambda^\sigma_\mu$ with
the strong field $F_{\mu\sigma}$. Since the product of two antisymmetric
tensors results in a symmetric tensor, the Ricci tensor has been reduced to
a symmetric tensor as desired. In order to avoid possible confusion, it
should be emphasised here that tensor properties are changed when the
Ricci tensor is interpreted in terms of physical quantities like strong
fields. As in the case of electromagnetic fields, the above geometric
formulation of strong fields is covariant only under the group of linear
transformations, so tensor properties respect only the linear group.
However, when the Ricci tensor is formed as a coupling of two strong
fields, it may give rise to an entirely different physical effect on a
particle moving in the coupled fields from possible effects the particle
may be influenced under a single strong field. Another important point that
should also be made clear is that the curvature tensor is considered to be
formed only when two strong fields are coupled together. It is assumed that
it is the coupling of two strong fields that gives rise to the curvedness
of spacetime structure that may be described by the field equations of
general relativity. In order to investigate the dynamics of a particle
under the influence of the physical affine Ricci tensor, defined in terms of
the affine connections which are now interpreted as strong fields, a metric
tensor can be introduced onto the spacetime manifold formed by the coupling
of two strong fields. The possible field equations that may govern the
metric tensor can be assumed to take the form of the field equations of
general relativity, since it has been shown that the field equations of
general relativity also admit a line element of Yukawa potential.\\
\\
\noindent
{\bf REFERENCES}\\
\\
Landau, L.D. and Lifshitz, E.M. (1975) {\it The Classical Theory of
Fields} (Pergamon Press).\\
Ho, Vu B (1994) {\it Spacetime geometry of electromagnetism} (preprint).\\
Ho, Vu B (1994) {\it Gravity as a coupling of two electromagnetic fields}
(preprint).

\end{document}